\documentclass[pre,twocolumn]{revtex4}

\usepackage[german,english]{babel}
\usepackage{amsmath}
\usepackage{mathrsfs}
\usepackage{graphicx}

\begin{document}

\title{L{\'e}vy flights as subordination process: first passage times}

\author{Igor M. Sokolov}
\affiliation{Institut f{\"u}r Physik, Humboldt Universit{\"a}t zu Berlin,
Newtonstra{\ss}e 15, 12489 Berlin, FRG}
\email{igor.sokolov@physik.hu-berlin.de}
\author{Ralf Metzler}
\affiliation{NORDITA -- Nordic Institute for Theoretical Physics,
Blegdamsvej 17, 2100 Copenhagen {\O}, Denmark}
\email{metz@nordita.dk}

\begin{abstract}
We obtain the first passage time density for a L{\'e}vy flight random
process from a subordination scheme. By this method, we infer the asymptotic
behavior directly from the Brownian solution and the Sparre Andersen
theorem, avoiding explicit reference to the fractional diffusion
equation. Our results corroborate recent findings for Markovian L{\'e}vy
flights and generalize to broad waiting times.
\end{abstract}

\maketitle

An important issue in the theory of stochastic processes is the problem of
first passage \cite{hughes,vankampen,feller,redner}: Its solution
is a key to the understanding of chemical reactions, stability of states of
dynamical systems under external perturbations, extinction of populations,
and many other problems in natural sciences. In normal Gaussian diffusion, a
standard method of solution of the first passage problems is the method of
images due to Kelvin \cite{redner,carslaw,feller}. However, for L\'{e}vy
flights (LFs) it has recently been demonstrated that the images method leads
to a result that contradicts the Sparre Andersen theorem, according to which
the first passage time density (FPTD) of a random walk process
asymptotically follows the universal $f(t)\sim t^{-3/2}$ behavior for any
symmetric distribution of jump lengths \cite{sparre,redner,chech_fpt}. This
statement is of great importance since it not only points out possible
inapplicability of Kelvin's method to superdiffusive jump processes, but
also poses intricate questions of a correct continuous limit for L\'{e}vy
flights and their description within fractional diffusion or Fokker-Planck
equations.

LFs are a paradigm for anomalous stochastic processes with a wide range of
applications such as chaotic dynamics, processes in plasma, transport in
micelles, or even quantum systems
\cite{bouchaud,report,zasla,yossinature,igorpt,chech_plas,walter}.
They can be defined in terms of a random walk process with long-tailed jump
length distribution $\lambda (x)\sim \sigma ^{\mu }/|x|^{1+\mu }$
($0<\mu <2$) \cite{klablushle}. Alternatively, LFs can be considered as 
appearing from a Langevin equation with $\delta$-correlated L{\'{e}}vy
noise \cite{west,fogedby,chechlong}. The characteristic function of an LF
given by  
\begin{equation}
P(k,t)=\exp \left( -K^{(\mu )}|k|^{\mu }t\right)   \label{lf_chf}
\end{equation}
is of stretched Gaussian type, where $P(k,t)\equiv \mathscr{F}\{P(x,t)\}=
\int_{-\infty }^{\infty
}P(x,t)\exp (ikx)dx$ is the Fourier transform of the probability density
function (PDF) $P(x,t)$ \cite{levy,hughes,bouchaud,yossinature}. In
what follows, the generalized diffusion constant $K^{(\mu )}$ will be set to
unity. In the limit $\mu =2$, 
the characteristic function (\ref{lf_chf}) reduces to the Gaussian 
$P(k,t)=\exp \left( -K^{(2)}k^{2}t\right) $, the characteristic function of a
standard random walk with Gaussian limit distribution and finite variance $%
\langle x^{2}(t)\rangle =2K^{(2)}t$ \cite{levy1,vankampen}, and finite higher
order moments. For the general case $0<\mu <2$, only fractional moments of
the form 
\begin{equation}
\langle |x|^{\delta }\rangle =\frac{2^{1+\delta /\mu }}{\mu \pi ^{1/2}}\frac
{\Gamma (1/2+\delta /2)\Gamma (-\delta /\mu )}{\Gamma (-\delta /2)}\Big(
K^{(\mu )}t\Big)^{\delta /\mu },
\end{equation}
exist with $\delta <\mu $ which can most easily be obtained by using
properties of Fox $H$-functions \cite{meno}. 

The fractional diffusion equation \cite{fogedby,report,meno,igorpt} 
\begin{equation}
\frac{\partial P(x,t)}{\partial t}=K^{(\mu )}\frac{\partial ^{\mu }}
{\partial |x|^{\mu }}P(x,t)  \label{fde}
\end{equation}
governing an LF reflects the inherent long jumps by the non-locality of the
fractional Riesz-Weyl operator \cite{chechlong,samko,report,meno} 
\begin{equation}
\frac{\partial ^{\mu }P(x,t)}{\partial |x|^{\mu }}\equiv \frac{1}{2\cos
\left( \pi \mu /2\right) \Gamma (2-\mu )}\frac{\partial ^{2}}{\partial x^{2}}
\int_{-\infty }^{\infty }\frac{P(\xi ,t)d\xi }{|x-\xi |^{\mu -1}},
\label{frw}
\end{equation}
for $1\le \alpha <2$, and an analogous expression for $0<\mu \le 1$; its
Fourier transform $\mathscr{F}\{\partial^{\mu}P(x,t)/\partial|x|^{\mu}\}$ is
$-|k|^{\mu}P(k,t)$ \cite{samko,chechlong}. In the limit $\mu
=2$, the fractional operator (\ref{frw}) reduces to the standard
second-order partial derivative, as it should.

In terms of the dynamical equation for the PDF, the
formulation of the first passage problem in terms of an absorbing
(Dirichlet) boundary condition requires a cutoff in the Riesz-Weyl integral 
(\ref{frw}) \cite{chech_fpt}. Contrasting the universal Sparre Anderson
result for the FPTD, it was found that the probability density of first
arrival at a site differs from the FPTD, and is explicitly dependent on the
index $\mu $ of the LF \cite{chech_fpt}. In the present work, we present an
alternative derivation of the FPTD for LFs based on a subordination to a
regular random walk process of the kind developed in Refs. \cite
{SubSok,igor_dirk}, leading us to a new, a priori unexpected twist: Although
the PDF of the free process (without boundaries) is described by the
fractional diffusion equation, Eq.(\ref{fde}), this equation does not
uniquely describe the whole process, i.e. the exact properties of its sample
paths. Depending on how anomalous transport statistics are introduced, we
will discuss the appropriateness or failure of the images method.

Let us consider a stochastic process subordinated to a discrete random walk
with some operational time that is defined by a one-sided L\'{e}vy law
\cite{SubSok}.
The notion of subordination implies that the corresponding random
process can be understood as follows \cite{feller}. The motion of the
random walker can be parameterized by the number of steps $n$, the PDF of the
random walker's position after $n$ steps being given by the PDF $P_{RW}(x,n)$. 
The corresponding characteristic function $\varphi _{n}(k)=\int_{-\infty
}^{\infty }P_{RW}(x,n)\exp \left( ikx\right) dx$ is determined by the
PDF of step lengths $\lambda (x)$ such that $\varphi
_{n}(k)=\left[ \lambda (k)\right] ^{n}$ in terms of the characteristic
function $\lambda (k)$ of the step length PDF, compare, for
instance, Ref. \cite{hughes}.

The number of steps $n$ itself that may be considered as the internal
operational time of the process, is a nondecaying random function of the
physical time (clock time) $t$. Denoting by $p_{n}(t)$ the probability to
perform exactly $n$ steps up to time $t$, we obtain 
\begin{equation}  \label{CTRW}
P(x,t)=\sum_{n}P_{RW}(x,n)p_{n}(t).
\end{equation}
At long times we suppose that a continuous operational time $T$ may be
introduced instead of the discrete index $n$. Moreover, we assume that
the continuous analog of $P_{RW}(x,n)$ exists. In this limit, we find from
(\ref{CTRW}) 
\begin{equation}  \label{Continuous}
P(x,t)=\int_{0}^{\infty }P_x(x,T)p_T(T,t)dT,
\end{equation}
where $P_x(x,T)$ is the PDF to be at $x$ at operational time $T$ and 
$p_T(T,t)$ is the PDF to be at operational time $T$ at clock time $t$. This
continuous limit corresponds exactly to the mathematical notion of the
subordination, see Sec. X.7 of Ref. \cite{feller}: `If $\{X(T)\}$ is a
Markov process with continuous transition probabilities and $\{T(t)\} $ a
process with non-negative independent increments, then $\{X(T(t))\}$ is said
to subordinate to $\{X(t)\}$ using the operational time $T$.'

Instead of having the PDF $P_{RW}(x,n)$ of a Brownian random walk, let us
assume the more general case that $P_{RW}(x,n)$ corresponds to an LF, so
that $P_{RW}(x,n)$ is given by a symmetric L\'{e}vy distribution of the form 
\cite{feller} 
\begin{equation}  \label{SyLe}
P_{RW}(x,n)=\left( l_{0}n^{1/\alpha }\right)^{-1}L\left(\frac{x}{l_0n^{1/
\alpha}},\alpha,0\right)
\end{equation}
($0<\alpha \leq 2$), where $l_0$ is a scaling factor with length as
physical unit. Note that the limit $\alpha=2$ corresponds to a Gaussian
profile for $P_ {RW}(x,n)$. Let us additionally assume that the number of
steps per unit of the physical (clock) time $t$ is distributed according to
some distribution with a power-law tail: $p(n,\Delta t=1)\propto n^{-1-\beta}
$, with $0<\beta\leq 1$. Then, according to the generalized central limit
theorem, at long times $p_{n}(t)$ tends to a continuous
limit distribution corresponding to the one-sided L{\'e}vy law 
\begin{equation}  \label{Oneside}
p_{T}(T,t)=\left(\frac{\tau_0}{t}\right)^{1/\beta}L\left( T\left(\frac{\tau_0}
{t}\right)^{1/\beta};\beta,-\beta\right).
\end{equation}
Here, $\tau_0$ is a scaling factor with physical unit time. The one-sided
character of (\ref{Oneside}) ensures that $p_T(T<0,t)\equiv 0$ \cite
{levy,feller}.

To obtain the limit distribution $P(x,t)$ based on relations
(\ref{SyLe}) and (\ref{Oneside}), we first Fourier-transform Eq.
(\ref{Continuous}) in respect to $x$. With expression (\ref{lf_chf}) for the
characteristic function of a symmetric L{\'e}vy stable density, we find 
\begin{eqnarray}
P(k,t)=\int_0^{\infty}&&\exp\Big(-|kl_0|^{\alpha}T\Big)\left(\frac{\tau_0}{t
}\right)^{1/\beta}  \nonumber \\
&&\times L\left(T\left(\frac{\tau_0}{t}\right)^{1/\beta};\beta,-\beta
\right)dT.  \label{Lapl}
\end{eqnarray}
This exactly corresponds to the Laplace-transform of a one-sided L\'{e}vy
density with Laplace variable $u=|k|^{\alpha}l_0^{\alpha}$. The Laplace
transform of a one-sided L{\'e}vy stable density is known \cite
{levy,hughes}: $\tilde{p}_T(u,t)=\exp\left(-u^{\beta}t/\tau_0
\right)$, and thus
\begin{equation}  \label{SBR}
P(k,t)=\exp\left(-|k|^{\alpha\beta}\left[l_0^{\alpha\beta}\tau_0^{-1}\right]
t\right).
\end{equation}
This expression, in turn, is the characteristic function of a symmetric
L\'{e}vy stable density with index $\alpha\beta$. The scaling factor 
$K^{(\alpha\beta)}\equiv l_0^{\alpha\beta}/\tau_0$ can be interpreted as the
associated fractional diffusion coefficient that depends on the indices of
the corresponding subprocesses only through their product $\gamma=\alpha\beta$.

The PDF (\ref{SBR}) fulfills the fractional diffusion equation (\ref{fde})
with order $\mu =\gamma $. This can also be shown on grounds of the
subordination scheme developed in Refs. \cite{SubSok,soklablu}. However, it
is remarkable that the dynamical equation (\ref{fde}) defines the PDF $P(x,t)
$, but not the process (or better, the identity of the subprocesses) itself;
unless we have to do with the limiting case $\gamma =2$ that necessarily
corresponds to $\alpha =2$ and $\beta =1$, defining the process uniquely. In
other words, although the PDFs of all processes with identical $\gamma $ are
the same, these processes may still differ in the fractal dimension of their
sample paths given by the set of jumps of random length corresponding to LFs
with index $\alpha $; and similarly they differ in the nature of the
connection between operational time $T$ and physical clock time $t$
according to the one-sided L{\'{e}}vy stable density with index $\beta $.
This issue is of great importance for first passage problems, as we
demonstrate now.

Our subordination procedure corresponds to a (random) change of the time
variable of the process from operational time $T$ to clock time $t$. This
allows us to solve a number of problems connected to the underlying random
process without explicitly referring to the fractional equation (\ref{fde}).
We here consider the first passage across a boundary located at $x=0$.

Let $\mathscr{S}_n(x_0)$ be the survival probability on the positive
semi-axis (i.e. the probability of not crossing the boundary within the
first $n$ steps) after starting at $x_0>0$ at $n=0$. According to the
Sparre Andersen theorem \cite{sparre,redner,feller}, the asymptotic form of
this probability does not depend on the jump length distribution if only it
is symmetric. For large number of steps, one invariably has $%
\Psi_n(x_0)\simeq c(x_0)n^{-1/2}$, where the prefactor $c(x_0)$ depends on
the initial position $x_0$, as well as on $\alpha$ and $l_0$. On subordinating
the number of steps $n$ to physical clock
time $t$ we see that the survival probability up to time $t$ is
\begin{equation}  \label{surv}
\mathscr{S}(t;x_0)=\sum_n\Psi_n(x_0)p_n(t),
\end{equation}
where $p_n(t)$ is the probability that exactly $n$ steps
occur within clock time $t$. Note that equation (\ref{surv}) corresponds to
the spatial average of Eq. (\ref{CTRW}). Changing from $n$ to the continuous
operational time variable $T$, we get: 
\begin{equation}
\mathscr{S}(t;x_0)=\int_{0}^{\infty }\Psi(T;x_0)p_T(T,t)dT.
\end{equation}
This and $\Psi_n(x_0)\simeq c(x_0)n^{-1/2}$ gives rise to
\begin{equation}
\mathscr{S}(t;x_0)
\simeq c(x_0)\frac{1}{\pi^{1/2}\beta}\Gamma\left(\frac{1}{2\beta}\right)
\left(\frac{\tau_0}{t}\right)^{1/(2\beta)},  \label{survi}
\end{equation}
as derived in the Appendix. We now obtain the FPTD,
\begin{equation}
f(t)=-\frac{d\mathscr{S}(t;x_0)}{dt}\simeq \frac{c(x_0)}{\pi^{1/2}\beta}
\Gamma\left(1+\frac{1}{2\beta}\right)\frac{\tau_0^{1/(2\beta)}}{t^{1+1/(2
\beta)}}.
\end{equation}
The following limiting cases can be distinguished:

(i) If the subordination from operational time $T$ to clock time $t$ through 
$p_T(T,t)$ is narrow with $\beta=1$, i.e., $p_T(T,t)=\delta(T-t/\tau_0)$, the
universal $f(t)\sim t^{-3/2}$ behavior according to the Sparre Anderson
theorem is recovered. In other words, in order to change this asymptotic
behavior, one has to consider $0<\beta<1$ explicitly.

(ii) If we consider the process subordinated to Gaussian diffusion ($\alpha=2
$), then $\beta=\gamma/2$ and 
\begin{equation}  \label{paradox}
f(t)\simeq \frac{2c(x_0)}{\pi^{1/2}\gamma}
\Gamma\left(1+\gamma\right)\frac{\tau_0^{1/\gamma}}{t^{1+1/\gamma}}.
\end{equation}
This result has the same scaling as the FPTD derived through the method of
images in Ref. \cite{chech_fpt}. A few words on the interpretation of this
seemingly paradox finding are in order. Result (\ref{paradox}) corresponds
to a random walk process with a Gaussian jump length density $\lambda(x)$,
so that the corresponding trajectory is that of a regular Brownian walk. It
is therefore perfectly legitimate to use the images method for such a
process, even though it is described by the fractional diffusion equation 
(\ref{fde}). In contrast, the genuine (`classical') LF discussed in Ref. \cite
{chech_fpt} corresponds to a broad $\lambda(x)\sim l_0^{\alpha}/|x|^{1
+\alpha}$ with $\alpha$ explicitly smaller than 2, but $\beta=1$. For this
strongly non-local process with fractal trajectory, the FPTD follows the
result (i). In this latter case, that is, the method of images fails and the
FPTD follows $f(t)\sim t^{-3/2}$.

(iii) In general, for a given $\gamma$ one has to make sure that the
inequalities $\gamma/2<\beta\leq 1$ are fulfilled, since simultaneously the
two conditions $\beta\leq 1$ and $\beta=\gamma/\alpha$ with $0<\alpha\leq 2$
have to be met. The FPTD for such a general $\gamma$ therefore shows the
following asymptotic behavior
\begin{equation}
f(t)\propto t^{-\delta}, \quad 3/2\leq \delta \leq 1+1/\gamma,
\end{equation}
where $0<\gamma\leq 2$. In this scheme, the Sparre Andersen decay with
exponent $3/2$ is the slowest one possible. This makes perfect sense since
due to the L{\'e}vy stable form of $p_n(t)\sim n^{-1-\beta}$ a broad
distribution of single jump events occurs in a finite time interval $(0,t)$,
increasing the likelihood of crossing the boundary within any given finite
time interval dramatically. The L{\'e}vy nature of $p_n(t)$ thus leads to an
oversampling of the space in comparison to a process $\beta=1$.

The subordination scheme developed here allows one to express the FPTD of a
random process solely on the basis of the properties of the Sparre Andersen
universality and the subordination map from operational time $T$
to physical clock time. Starting off from an LF in $(n,t)$
coordinates, we introduce a broad distribution $p_n(t)\sim n^{-1-\beta}$ of
events per clock time interval $\Delta t$. We find that the resulting
process is governed by the fractional diffusion equation (\ref{fde}) whose
order $\gamma=\alpha\beta$ is a product of the L{\'e}vy index of the jump
length distribution, $\alpha$, and the subordination distribution $p_n(t)$, 
$\beta$. The knowledge of $\gamma$ alone is therefore insufficient to
deduce the exact form of the jump length PDF and the trajectory it
gives rise to.

As a direct consequence, the resulting FPTD in the limit $p_n=\delta_{n,1}$
(the trivial subordination with $\beta=1$) fulfills the Sparre Anderson
universality for any process with symmetric jump length distribution. This
case includes the case of genuine LFs as those discussed in Ref. \cite
{chech_fpt}, and is violated by the method of images for all $0<\alpha <2$.
Conversely, for $\alpha=2$, the subordination process has the trajectory of
a normal Brownian random walk and is amenable to the images method
to determine the FPTD. The result scales like $f(t)\sim t^{-1-1/\gamma}$ as
previously obtained. In the case of general $\alpha$ and $\beta$, the range
spanned by the exponent in the FPTD $f(t)\sim t^{-1-\delta}$ is $3/2\le
\delta \le 1+1/\gamma$.

We believe that above findings help in interpreting the inadequacy of the
images method for genuine LFs as found in \cite{chech_fpt} and moreover show
that caution is necessary when generalizing well-known results from ordinary
to anomalous diffusive processes.

We acknowledge helpful discussions with Aleksei Chechkin and Yossi Klafter.

\begin{appendix}

\section{Calculation of the ($-1/2$)-order moment}

To obtain the final form of Eq. (\ref{survi}), we have to evaluate the
integral
\[
\int_0^{\infty}T^{-1/2}\left(\frac{\tau_0}{t}\right)^{1/(2\beta)}L\left(T
\left(\frac{\tau_0}{t}\right)^{1/\beta};\beta;-\beta\right),
\]
which is equal to the ($-1/2$)-order moment
\[
M_{-1/2}(\beta)=\int_0^{\infty}\xi^{-1/2}L(\xi;\beta;-\beta)d\xi
\]
of the one-sided L{\'e}vy stable density $L(\xi;\beta;-\beta)$. To calculate
this expression, one can follow two alternative routes. As such techniques
are useful when dealing with L{\'e}vy processes, we sketch both ways.

(i) The density $L(\xi;\beta;-\beta)$ is the inverse Fourier transform of the
characteristic function $\varphi(k)=\exp\left(-|k|^\beta e^{-i\pi\beta/2}
\right)$ for $k>0$, continued to the negative semi-axis by $\varphi(-k)=
\varphi^*(k)$. Then,
\[
M_{-1/2}(\beta)=\int_0^\infty \frac{d\xi}{\xi^{1/2}}\mbox{Re}\int_0^
\infty\frac{dk}{\pi} e^{-ik\xi}\exp\left(-|k|^\beta e^{-i\pi\beta/2} \right).
\]
Interchanging both integrals, one can express the integral over $\xi$ through
Fresnel integrals to obtain $\int_0^\infty \xi^{-1/2}\exp(-ik\xi)d\xi=\sqrt{
\pi/2}(1-i)k^{-1/2}$. Changing then $-ik$ to $u$ we find for the overall
integral:
\[
M_{-1/2}(\beta)=\pi^{-1/2}\mbox{Re}\int_0^{-i\infty}du\exp\left(
-u^\beta\right))u^{-1/2}.
\]
Noting that the integrand is analytical everywhere except for $u=0$, vanishes
for $|u|\rightarrow\infty$ in the lower right quadrant of the complex plane,
and that the integral converges, we can change the contour of integration
to the positive real axis, so that
\begin{equation}
M_{-1/2}(\beta)=\int_0^\infty du \exp\left(-u^\beta\right)u^{-1/2}.
\end{equation}
Substituting $\eta=u^\beta$, we recover the
the integral representation of the $\Gamma$-function, so
that Eq. (\ref{survi}) follows.

(ii) Alternatively, the density $L(\xi;\beta;-\beta)$ is the inverse Laplace
transform of $\exp\left(-u^\beta\right)=\beta^{-1}H^{1,0}_{0,1}\left[u\left|
\begin{array}{l}-;-\\(0,1/\beta)\end{array}\right.\right]$ expressed in terms
of the Fox $H$-function \cite{srivastava,meno}.
By standard methods \cite{srivastava,gloeno}, the inverse Laplace
transform can be performed, yielding
\[
L(\xi;\beta;-\beta)=(\beta\xi)^{-1}H^{1,0}_{1,1}\left[\frac{1}{\xi}\left|
\begin{array}{l}(0,1)\\(0,1/\beta)\end{array}\right.\right].
\]
After a substitution, the resulting integral
\[
M_{-1/2}(\beta)=\frac{1}{\beta}\int_0^{\infty}z^{-1/2}H^{1,0}_{1,1}
\left[z\left|\begin{array}{l}(0,1)\\(0,1/\beta)\end{array}\right.\right]
\]
can be easily evaluated, by noting that it corresponds to the Mellin
transform $\hat{g}(s)\equiv\int_0^{\infty}t^{s-1}g(t)$ of $H^{1,0}_{1,1}(z)$
at $s=1/2$. The result can be directly identified with the definition of
the $H$-function \cite{srivastava,meno}, so that $M_{-1/2}(\beta)=\beta^{-1}
\Gamma(1/[2\beta])/\Gamma(1/2)$, reproducing Eq. (\ref{survi}).

\end{appendix}

\end{document}